\documentclass[conference]{IEEEtran}
\IEEEoverridecommandlockouts
\usepackage{cite}
\usepackage{amsmath,amssymb,amsfonts}
\usepackage{algorithmic}
\usepackage{graphicx}
\usepackage{textcomp}
\usepackage{xcolor}
\usepackage{subfigure}
\def\BibTeX{{\rm B\kern-.05em{\sc i\kern-.025em b}\kern-.08em
    T\kern-.1667em\lower.7ex\hbox{E}\kern-.125emX}}
    
\usepackage{caption}

\begin{document}

\title{INTCP: Information-centric TCP for Satellite Network}

\author{
\thanks{This work is partly supported by the National Key R$\&$D program of China (2019YFB1802701), Guangdong Basic and Applied Basic Research Foundation (2019B1515120031), NSFC grant (61872213, 62032013, 61432009). corresponding author: zhangxg@pku.edu.cn}

\IEEEauthorblockN{1\textsuperscript{st} Jinyu Yin}
\IEEEauthorblockA{\textit{Wangxuan Institute of Computer Technology} \\
\textit{Peking University}\\
yinjinyu@pku.edu.cn}
\and
\IEEEauthorblockN{2\textsuperscript{nd} Li Jiang}
\IEEEauthorblockA{\textit{Wangxuan Institute of Computer Technology} \\
\textit{Peking University}\\
jl99888@pku.edu.cn}
\and
\IEEEauthorblockN{3\textsuperscript{rd} Xinggong Zhang}
\IEEEauthorblockA{\textit{Wangxuan Institute of Computer Technology} \\
\textit{Peking University}\\
zhangxg@pku.edu.cn}
\and
\IEEEauthorblockN{4\textsuperscript{th} Bin Liu}
\IEEEauthorblockA{\textit{Department of Computer Science and Technology} \\
\textit{Tsinghua University}\\
liub@tsinghua.edu.cn}
}


\maketitle

\begin{abstract}
Satellite networks are booming to provide high-speed and low latency Internet access, but the transport layer becomes one of the main obstacles. Legacy end-to-end TCP is designed for terrestrial networks, not suitable for error-prone, propagation delay varying, and intermittent satellite links. It is necessary to make a clean-slate design for the satellite transport layer. 
This paper introduces a novel Information-centric Hop-by-Hop transport layer design, INTCP. It carries out hop-by-hop packets retransmission and hop-by-hop congestion control with the help of cache and request-response model. Hop-by-hop retransmission recovers lost packets on hop, reduces retransmission delay. INTCP controls traffic and congestion also by hop. Each hop tries its best to maximize its bandwidth utilization and improves end-to-end throughput. The capability of caching enables asynchronous multicast in transport layer. This would save precious spectrum resources in the satellite network.
The performance of INTCP is evaluated with the simulated Starlink constellation. Long-distance communication with more than 1000km is carried out. The results demonstrate that, for the unicast scenario INTCP could reduce $42\%$ one-way delay, $53\%$ delay jitters, and improve $60\%$ throughput compared with the legacy TCP. In multicast scenario, INTCP could achieve more than $6X$ throughput.   
\end{abstract}



\section{Introduction}
In recent years, satellite networks are booming again due to new technologies such as satellite miniaturization and rocket reusability~\cite{SpaceX}. Amazon Blue Origin plans to lunch $3296$ satellites on the orbits of 590/610/630km~\cite{Yani}, while Starlink will lunch $11943$ satellites on the 3 orbits rings 340/550/1150km~\cite{SpaceX}. In 2030, there will be more than 25000 satellites in space~\cite{Yani}. Most of them are Low-earth-orbit (LEO) and aims to provide ``High speed, Low latency broadband connectivity across the globe'' (Starlink's slogan), especially for the rural area, developing countries, aviation, and marine communication.

It is still a question for the satellite network to provide comparable bandwidth and latency with terrestrial networks.
The end-to-end transport layer plays a critical role. It guarantees congestion avoidance and reliable end-to-end transmission.
But the widely deployed transport layer of the TCP model~\cite{JSAC} is designed for terrestrial networks. It controls congestion windows by the premise that the network congestion results in packet loss.
In satellite networks with a relatively high channel error and dynamic topology, packet loss occurs not only by congestion but also by physical layer errors. This would confuse the loss-based congestion control model.
Secondly, the large bandwidth-delay product of satellite networks also deteriorates the performance of TCP~\cite{TCPloss}. In the Starlink constellation working as Bent-Tube, the Round-Trip-Time (RTT) is about $40-50ms$ tested by users' equipment~\cite{SpaceX}. If you connect two locations more than $1000km$, the RTT will be increased proportionally. Larger RTT will slow down the increment of TCP windows~\cite{JSAC}.
Thirdly, relative movement among satellites and ground stations makes paths broken frequently. LEO satellites fly over one's head for just a few minutes~\cite{Mark18}. Ground-Satellite-Link (GSL) handover occurs frequently. 
Besides, Inter-satellite-link (ISL) setup time may range from a few seconds to tens of seconds~\cite{Conext19}, depending on the relative velocity and power of satellites.  
Intermittent paths make it hard to maintain a reliable end-to-end connection for the current TCP protocol.    

To this end, several new transport layer designs such as TCP Hybla, TCP Spoofing, and TCP splitting, have been proposed to improve the throughput on satellite links~\cite{Hybla,TCPEx, DSTP, Spoofing, Splitting, PEP}. 
TCP Hybla~\cite{Hybla} is a modified version of TCP Reno with a specific design considering a large RTT of satellite networks. But it is still loss-based congestion control that ignores packet loss by channel errors. 
Besides, another specific Splitting-TCP architecture, such as TCP Spoofing~\cite{Spoofing}, TCP splitting~\cite{Splitting}, Performance Enhancing Proxy (PEP)~\cite{PEP}, has been carried out to take advantage of specific TCP version for satellite links without changing the protocol stack of end-users. Splitting-TCP breaks off end-to-end TCP connections and establishes segmented transport protocols by terrestrial part and the satellite part respectively. The satellite part optimizes its own TCP to efficiently deliver data over satellite networks, while the terrestrial part uses legacy TCP which is transparent to end-users. Unfortunately, it falls short of optimizing transmission on satellite links.

Hop-by-hop transport layer design is considered more suitable for the LEO network due to the fine-grained congestion control. RCP~\cite{RCP} calculates proper sending rate on each router and feedbacks the minimal sending rate over the path to sender, which is used to limit the actual sending rate. But the control loop of RCP is still end-to-end, so it's not agile enough to transient congestion and bandwidth fluctuation. Another kind of hop-by-hop design~\cite{SIGCCR92} uses back-pressure algorithm to determine whether the sending rate should be limited per router. Such design is based on leveraging buffer occupancy of the bottleneck or the available bandwidth of links, but the acquisition of these parameters are not inherently supported in network, especially in wireless satellite network. In this paper, we propose a universal hop-by-hop scheme for congestion control in LEO network.

\begin{figure}[!t]
\begin{center}
\includegraphics[width=0.9\linewidth]{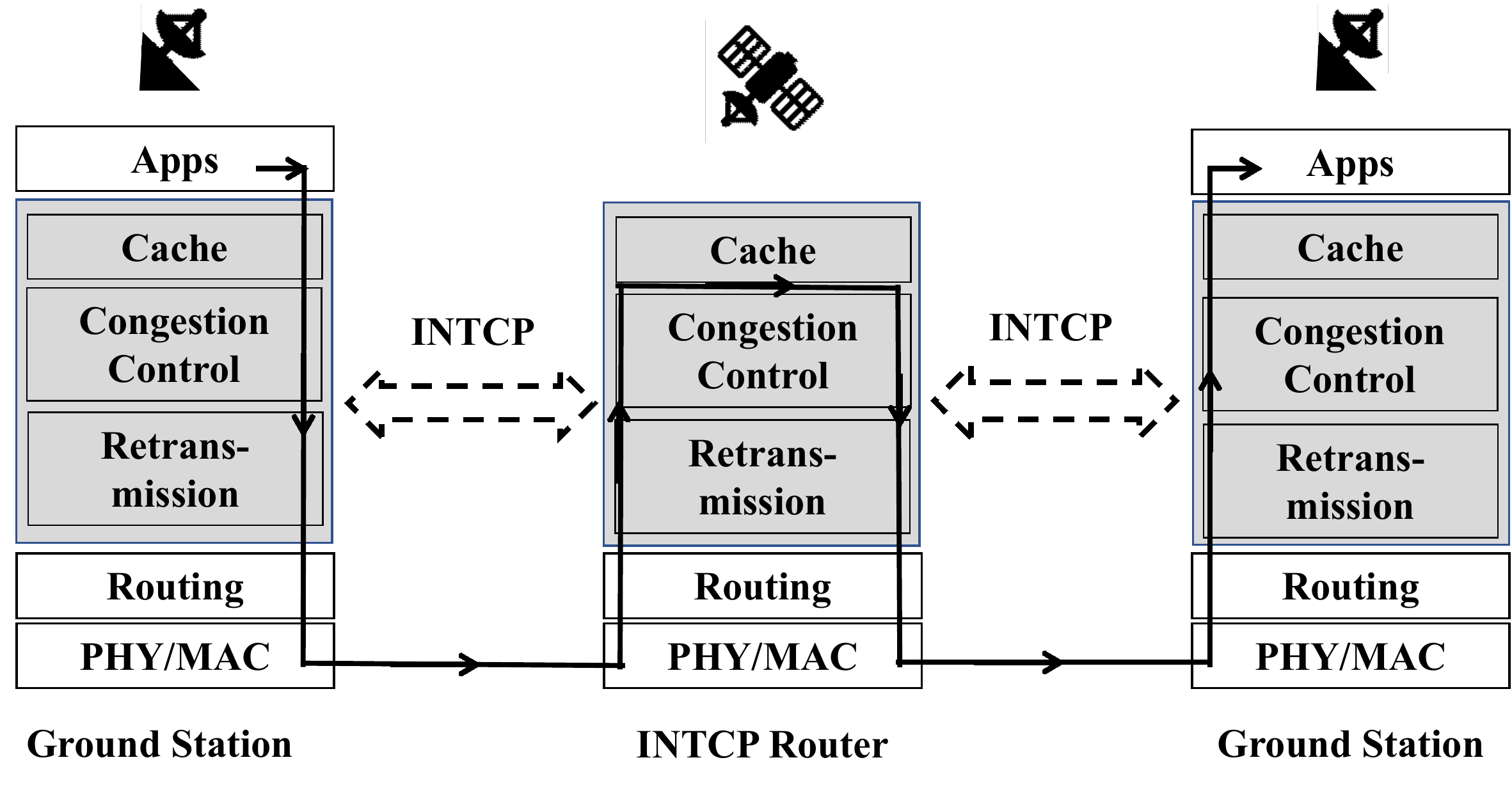} 
\end{center}
\caption{Paradigm of Information-centric TCP (INTCP) for Satellite Networks. INTCP stack with grey color is equipped in ground station and relaying satellite routers, and the transport layer is segmented hop-by-hop.}
\label{fig_arch}
\end{figure}

In this work, we are inspired by the idea of splitting-TCP and show that Information-centric TCP (INTCP) model can improve end-to-end transport performance over error-prone satellite links. The architecture of INTCP is illustrated in Fig.~\ref{fig_arch} with three modules: \emph{Cache, Congestion Control, and Retransmission}. It splits end-to-end path into segments, and provides hop-by-hop reliable transmission and congestion control. INTCP recovers packet loss on hop, instead of end-to-end packet retransmission in legacy TCP. This significantly reduces the delay of packet recovery, especially for satellite links with large RTT. Hop-by-hop Congestion  Control maximizes the link utilization on each hop. This would exploit the capacity of satellite links while avoiding the impairment of bottleneck link. Furthermore, network-embedded Cache copes with fan-in and fan-out traffic, and cache packets for multicast and hop-retransmission.   
 

INTCP keeps the three fundamental designs of Information-Centric Network (ICN)~\cite{Networkingnamedcontent, Asurveyofinformation} in transport layer, but ignores others. 1) \textbf{Named data}. Each packet is identified by the \textit{name} which is used for request and response, not for routing. 2) \textbf{Request-Response model}. A Requester would firstly issue an INTEREST message. Then the Responder responds with DATA message. 3) \textbf{Cache}. All incoming packets would be cached in local storage. Distinguished from the previous ICN and Name-data Network (NDN), INTCP is not designed for narrow waist of network layer, but for transport layer of TCP. It is compatible with any routing schemes, such as IP, Identity, or geographic routing.

With the aforementioned elaborate designs, INTCP is able to tackle the challenges faced by satellite networks. Firstly, with \textit{Hop-by-hop Congestion Control}, Responder on each hop can react to link condition changes quickly, and improve bandwidth utilization by hop-level other than end-to-end level congestion control. Furthermore, with the help of cache, INTCP can improve end-to-end throughput by \textit{Link-Multiplexing}. Each hop works separately and tries its best to send more data to the next hop. Due to time-varying capacity among satellite hops, extra incoming data would be stored in Cache, and be delivered to the next-hop when the link conditions become better.  Secondly, INTCP enables \textit{Hop-by-Hop Loss Recovery}. Once a packet is lost on a hop, Requester just re-issues a new INTEREST message and Responder retransmits the lost packets. The recovery delay is only one-hop instead of end-to-end RTT in the TCP model. Lastly, INTCP provides the instinctive capability of \textit{Multicast}. Thanks to the designs of cache and Request-Response model, any node can request the packets stored in the cache and provides asynchronous multicast in satellite network with precious spectrum resource. 

To evaluate the performance of INTCP, we carried out extensive simulated experiments with the trace of Starlink Constellation. It consists of $1160$ satellites on a $1000$km low-earth orbit. For random two terminals on Earth, they connect through moving satellites. On average, INTCP is able to improve end-to-end throughput by about $60\%$ and reduce one-way delay by around $42\%$ compared to Reno and Hybla. For multicast scenarios, INTCP is able to save $20\%$ bandwidth consumption. The evaluation results show that INTCP significantly improves the performance of the satellite network. It is a feasible and new paradigm for satellite transport layer.

The rest parts of the paper are organized as follows. The architecture of INTCP is elaborated in Sec.~\ref{sec_router}. The Segmented Congestion Control and Retransmission Mechanism are introduced in Sec.~\ref{sec_segcc} and ~\ref{sec_retrans}. Sec.~\ref{sec_eva} demonstrates the performance gains by extensive experiments. Finally, we summarize the paper in Sec.~\ref{sec_conclusion}.

\section{Information-centric Transport Layer}
\label{sec_router}
\begin{figure}[!t]
\begin{center}
\includegraphics[width=0.9\linewidth]{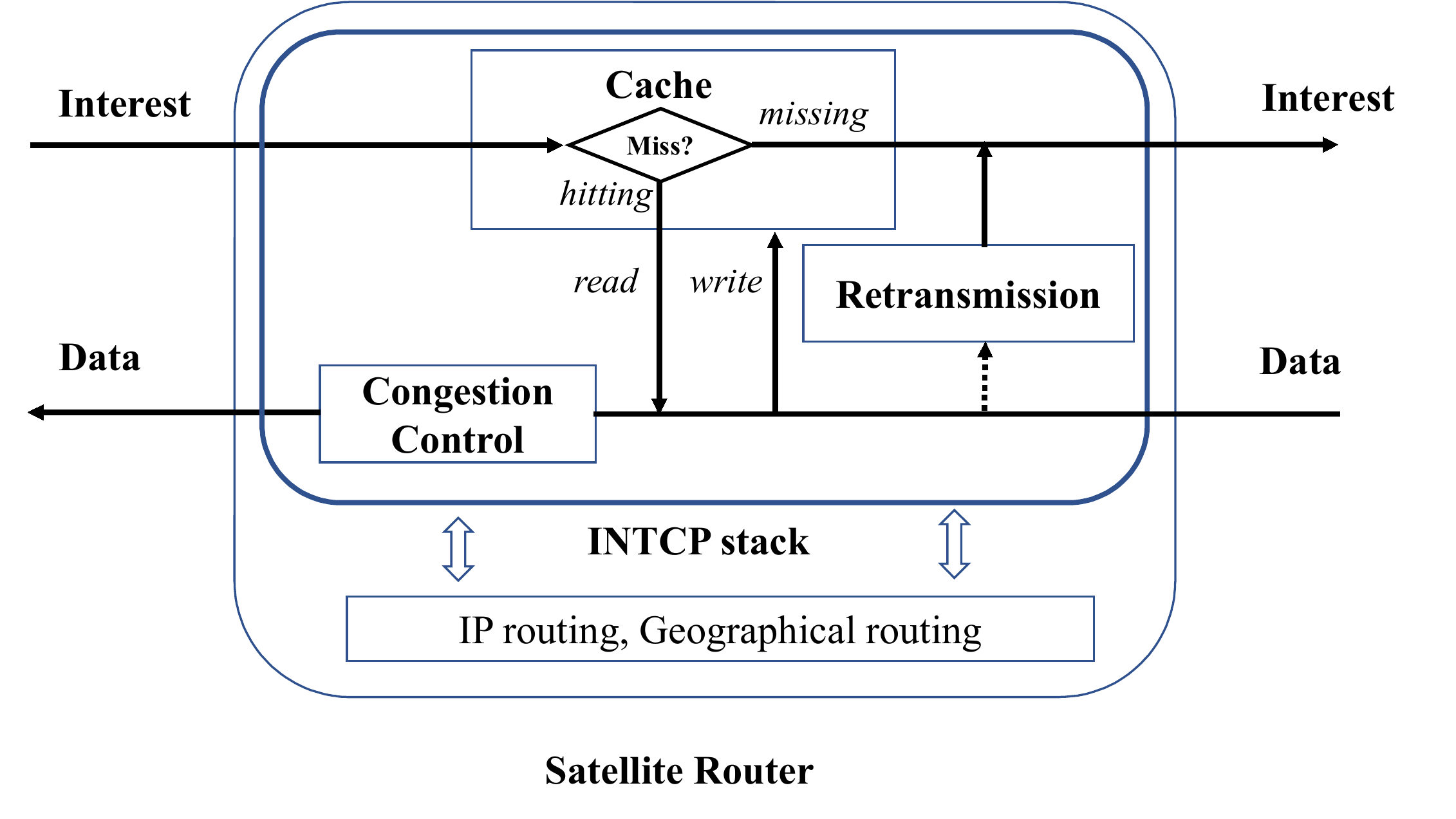} 
\end{center}
\caption{Architecture of Satellite Router} 
\label{fig_router}
\end{figure}

INTCP stack, the information-centric transport layer, is equipped on ground stations and INTCP routers. It operates congestion control and retransmits lost packets on each hop. INTCP mainly consists of three modules: Cache, Congestion Control, and Retransmission, as shown in Fig.~\ref{fig_router}. The rest layers, such as physical layer, link layer, network layer, and application layer are kept as same as in TCP/IP architecture. This provides a viable way to be compatible with current Internet devices and applications.

The transmission paradigm of INTCP is \textit{Request-Response-Cache}. When an INTEREST message is coming, Responder first tries to retrieve data in Cache. If found, the data will be responded to Requester. If not, the INTEREST message will be forwarded to the next hop. When a DATA message is incoming, the data will be stored in Cache for further retransmission and multicasting. At the same time, The data in Cache will be responded to Requester.

On one hop, Requester runs the congestion control algorithm based on the network state of this hop, such as packet loss, RTT, available bandwidth, etc, then periodically sends the controlling result, i.e. data sending rate, to Responder of this hop. Responder adjusts sending rate to avoid congestion on the hop. Besides, it retransmits lost packets indicated by the interests from Requester. The lost packets are stored in Cache and can be retransmitted immediately. It avoids the long latency of end-to-end packet loss recovery in legacy TCP.     

\textbf{INTEREST message}: The format of INTEREST message is like: \textit{(Requester, Responder, Name, RangeStart, RangeEnd, SendRate)}. The \textit{Name} indicates the name of content or service. The data of content or service is further segmented into packets. Each data packet is labeled by \textit{(RangeStart, RangeEnd)}, representing the byte-level range of the data in the whole content. And in an INTEREST message, the fields represent the range of data it requests for.

The expected send rate for the Responder of a hop, \textit{SendRate}, is piggybacked in INTEREST message. Once received a packet, Requester records the packet's delay, arriving rate, and losses, then figures out a send rate according to these pieces of information and feedback it to Responder for congestion control. 

Since the Request-Response model is stateless, Responder doesn't know who requests data.
The field of \textit{Requester} indicates the owner of INTEREST message. When network layer uses IP protocol, it is the IP address of Requester. The Responder responds interest by sending data to this address.
The field \textit{Responder} records the address of the Responder. The satellite router forwards interest to Responder by this address.

\textbf{DATA Message}: DATA message is replied from Responder to Requester. The format of DATA message is like: \textit{(Requester, Responder, Name, RangeStart, RangeEnd, Timestamp, Data)}. The field of \textit{Name, RangeStart, RangeEnd, Requester, Responder} has been discussed in the above paragraph. \textit{Timestamp} field indicates the emission time of packet on Responder. Requester uses it to calculate the one-way delay of data transmission on this hop.
The field of \textit{Data} comprises the payload.

\section{Hop-by-Hop Congestion Control}
\label{sec_segcc}

\begin{figure}[!t]
\begin{center}
\includegraphics[width=0.9\linewidth]{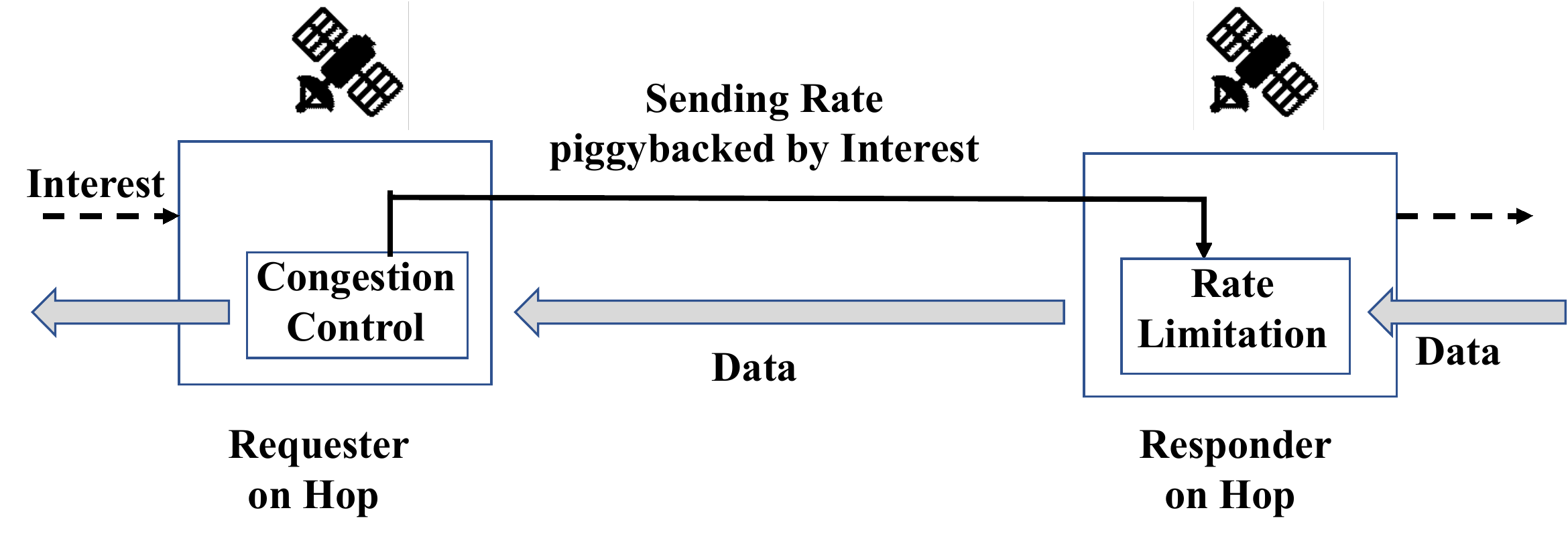} 
\end{center}
\caption{Hop-by-hop Congestion Control.} 
\label{fig_seg_cc}
\end{figure}

In INTCP, congestion control is segmented by hop as shown in Fig.~\ref{fig_seg_cc}. Each hop independently performs congestion control. At one end of hop, Requester detects network state and runs congestion control algorithm to get a proper data sending rate, periodically feeding it back to Responder. At the other end of hop, Responder is responsible for controlling DATA message sending to avoid congestion and achieve bandwidth fairness on the hop. 

Networks states of hop include one-way-delay (OWD), packet loss and available bandwidth. To get OWD, Responder write the \textit{Timestamp} into data packet head when sending it, then the Requester calculates the difference between the time it receives this packet and \textit{Timestamp} in this packet as OWD. packet loss and history throughput can be calculated straightforward on Requester side. This information implicitly stands for the hop's network conditions and congestion status. Besides, Requester also takes the available size of buffer into consideration. As multiple flows could exist simultaneously on one hop, in order to solve the fairness problem between flows, the congestion control module of satellite router gathers information of all the flows and uniformly calculates available bandwidth for them, thus a fair bandwidth allocating state can be easily reached.

Once the data sending rate piggybacked by interest is parsed by Responder, the Responder will use this value to control the data sending process, no matter where the data is from(from the previous hop or from its cache). We use the token bucket algorithm to realize the rate control function.

Specifically, the goal of our congestion control algorithm is to achieve fairness, high bandwidth utilization and low queuing delay on the hop. It also avoids buffer overflow on Requester. When the buffer on Requester is nearly full and the predicted output bandwidth is less, Requester would generate a smaller sending rate, then the Responder lowers its sending rate correspondingly. If the responder is forwarding data from previous hop, the buffer on it will also be filled up, since the input rate is 
higher than output rate. At this time, the responder, as a requester on the previous hop, will also generate a smaller sending rate. In this way, when the bottleneck of the network is about to have congestion, the upstream hops will lower their sending rate soon one by one, avoiding too much data to be sent to the bottleneck, which leads to queuing delay increasing and packet loss. While the legacy end-to-end loss-based or RTT-based congestion control algorithms are not suitable for time-varying satellite links, our hop-by-hop algorithm is robust to the variation. The conditions of segmented network can be detected more preciously compared to end-to-end network, and the feedback delay is less than the legacy end-to-end TCP. So INTCP is more agile in satellite network. Besides, different congestion control algorithms can coexist on multiple hops at the same time. One hop with terrestrial link can use the regular methods, like Cubic or Reno. Other hops with satellite links can use our rate-based congestion control. The buffer on satellite router makes it possible to match different rates of hops. Thus, the total bandwidth utilization will be improved.

\section{Retransmission Mechanism}
\label{sec_retrans}
\begin{figure}[!t]
\begin{center}
\includegraphics[width=0.9\linewidth]{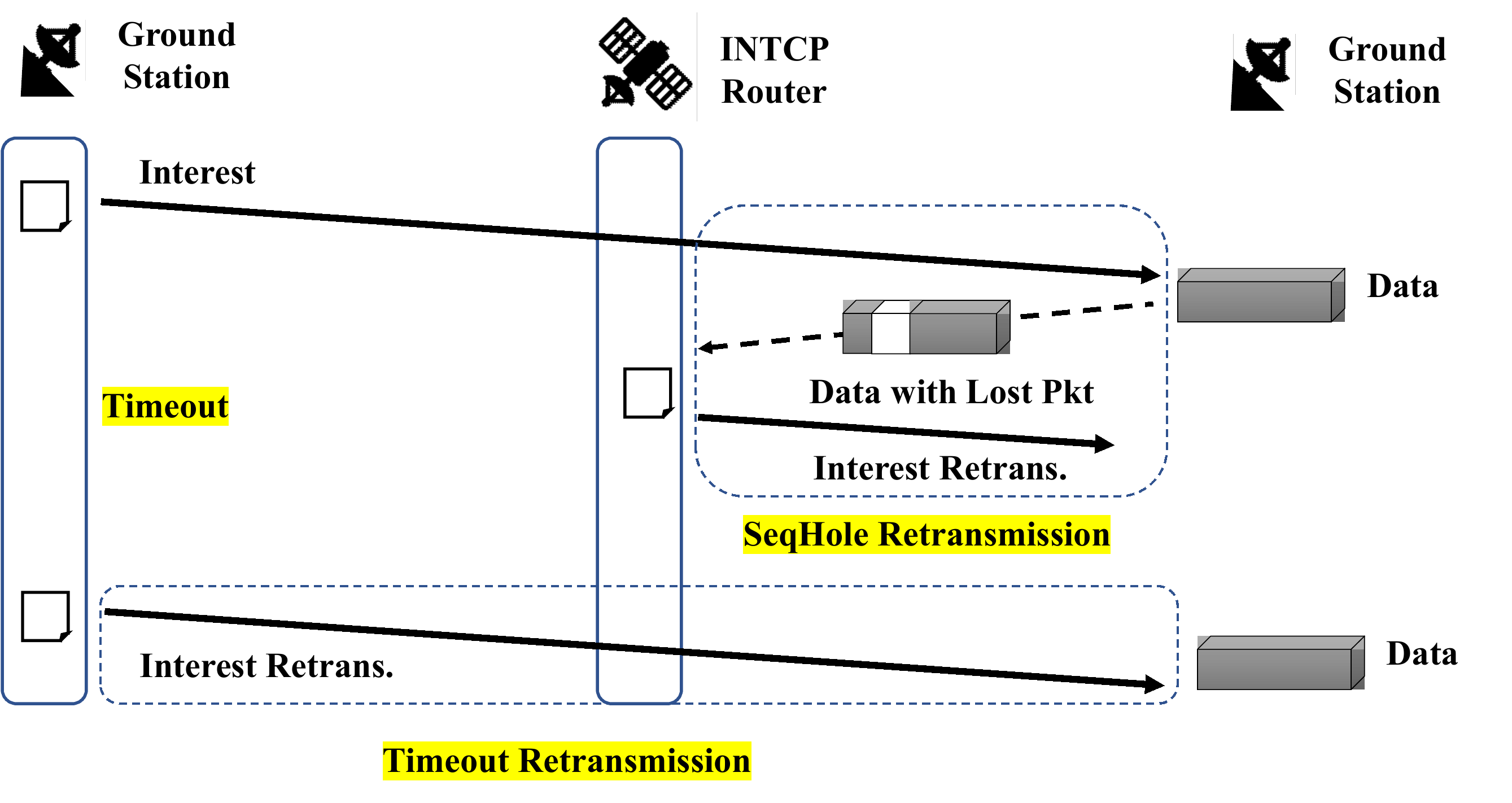} 
\end{center}
\caption{Hop-by-hop Retransmission} 
\label{fig_retrans}
\end{figure}

INTCP uses a Selective Retransmission mechanism to guarantee reliable data transport in satellite networks. When receiver found packet losses, it would reissue one retransmission Interest message to request the lost packets. In INTCP, there are two retransmission mechanisms: SeqHole Retransmission and Timeout Retransmission, as shown in Fig.~\ref{fig_retrans}. 

\textbf{SeqHole Retransmission} is belong to hop-by-hop retransmission. It detects and recovers lost packets on each hop. When Requester of hop receives a DATA message, it checks whether the packet number is sequential according to the field of \textit{RangeStart} and \textit{RangeEnd}. Once a range of data has been skipped by three subsequent packets, the missed range of data is assumed to be lost. Requester will issue a SeqHole Retransmission Interest message with the format:
\begin{equation}
    \textbf{INTEREST} \textit{(name, LostFrom, LostEnd, TTL)}
\end{equation}

The fields of $(LostFrom, LostEnd)$ indicate the retransmission range of lost packets. Responder sends back Data message by this range.

The field TTL represents how many hops that this message could be forwarded. Usually, the TTL of SeqHole Retransmission message is $1$. It means that the retransmission Interest can only be forwarded within one hop. This aims to avoid repeated data requests since all downstream nodes would notice packet losses and emit Retransmission Interest. The TTL of one hop makes a constraint that the retransmission requests are not repeated.

\textbf{Timeout Retransmission} is an end-to-end retransmission mechanism. When Consumer issues an INTEREST message, it will record the \textit{name, range, time} in the local Retransmission Timeout table (RTO). When receiving a packet, Consumer removes the records from RTO table. If RTO runs out and the data are yet not received, Consumer would issue a Timeout Retransmission Interest message.

RTO Retransmission is complementary to the hop-by-hop retransmission. The hop-by-hop retransmission enables on-path packet recovery. This would significantly reduce the recover delay. But it can not guarantee end-to-end reliable transmission since it recovers loss packet within one hop. When satellite orbit changes or link is broken, the lost packet not be recovered by hop should be retransmitted from server by RTO retransmission mechanism. 

\begin{figure*}[!t]
\centering
\subfigure[throughput under different bandwidth fluctuation period]{
\begin{minipage}[t]{0.3\linewidth}
\centerline{\includegraphics[width=1\linewidth]{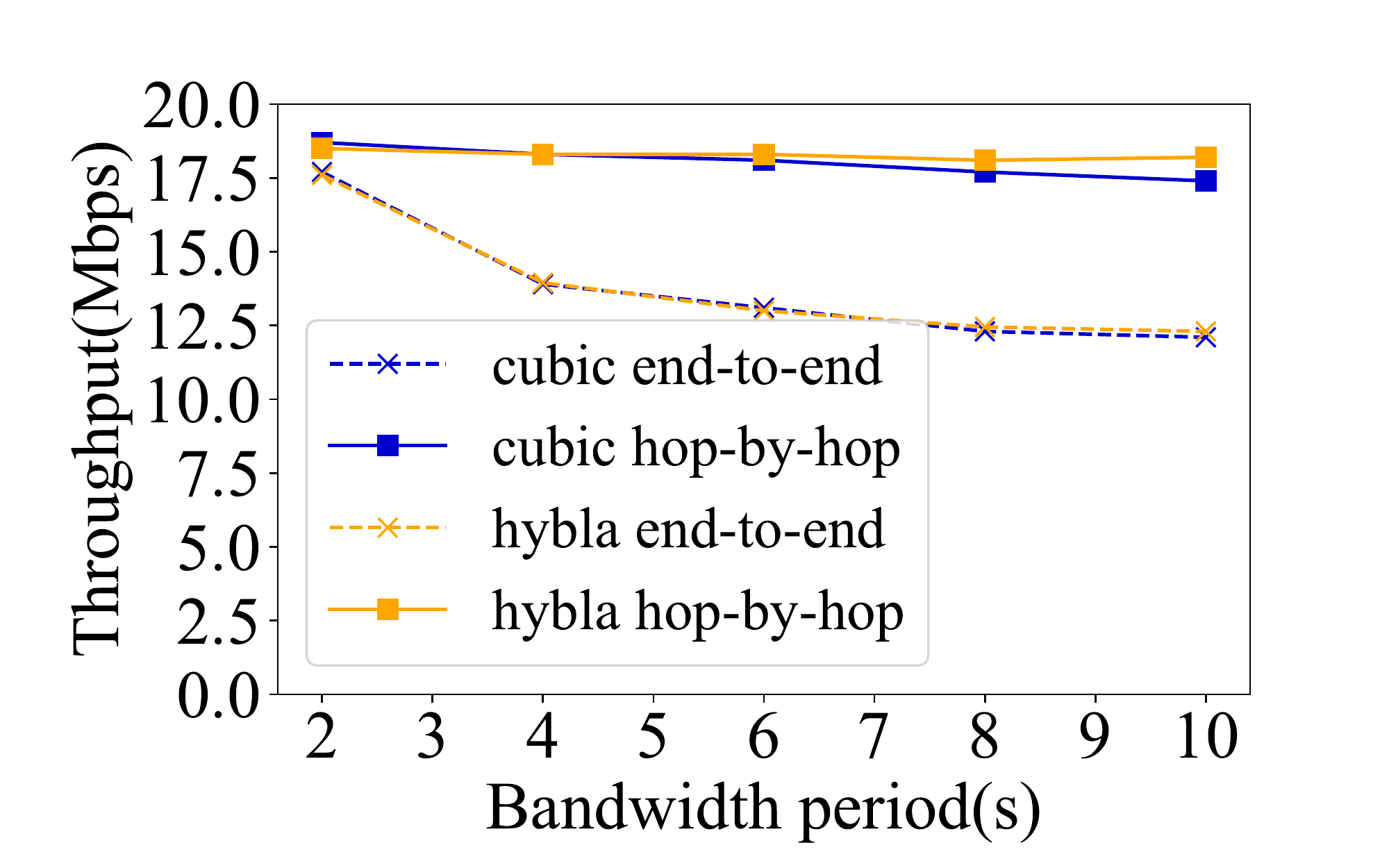}}
\label{fig_varIntv}
\end{minipage}
}
\subfigure[throughput under different intermittent time]{
\begin{minipage}[t]{0.3\linewidth}
\centerline{\includegraphics[width=1\linewidth]{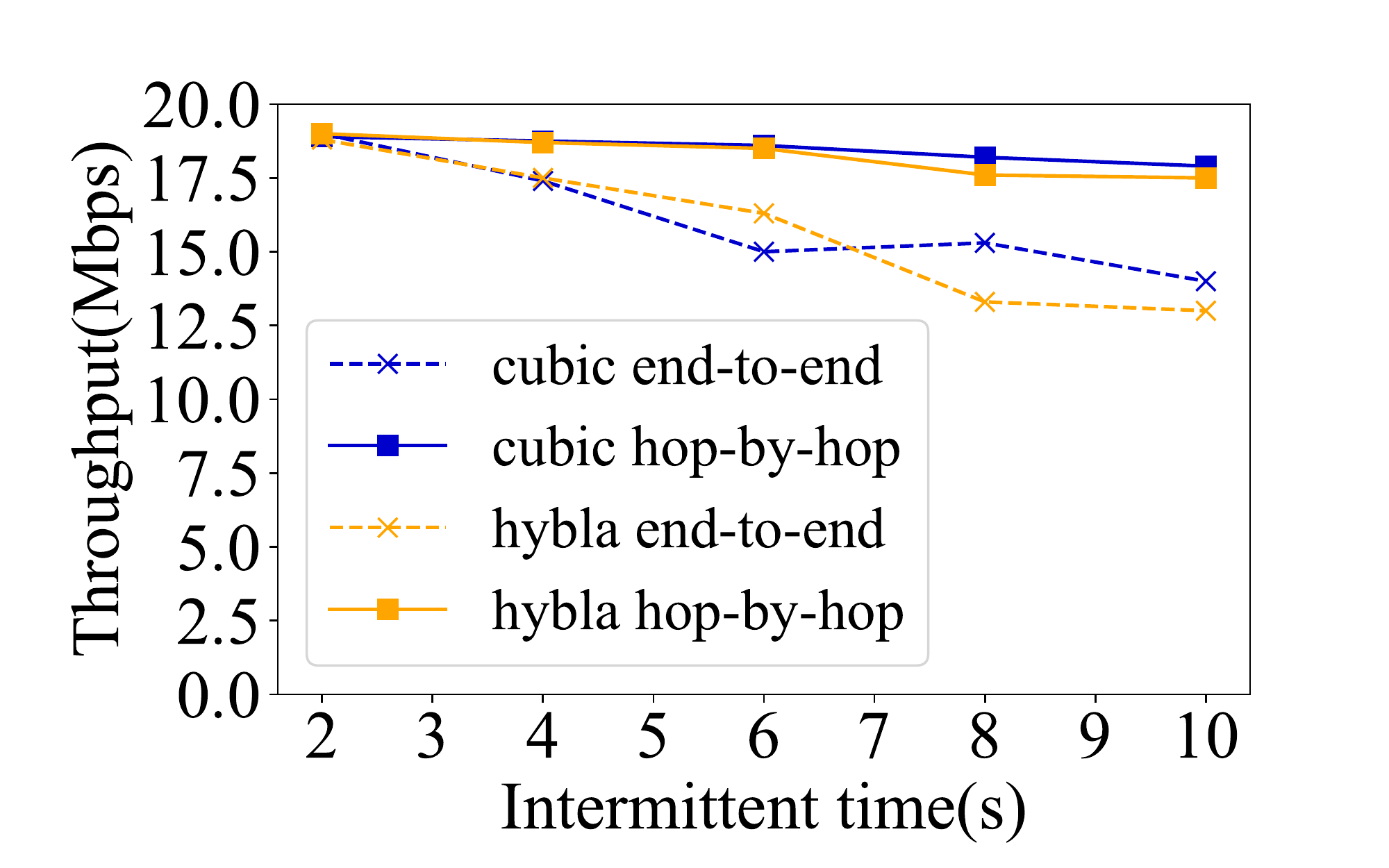}}
\label{fig_itmDown}
\end{minipage}
}
\subfigure[one-way delay distribution under different satellite link RTT]{
\begin{minipage}[t]{0.3\linewidth}
\centerline{\includegraphics[width=1\linewidth]{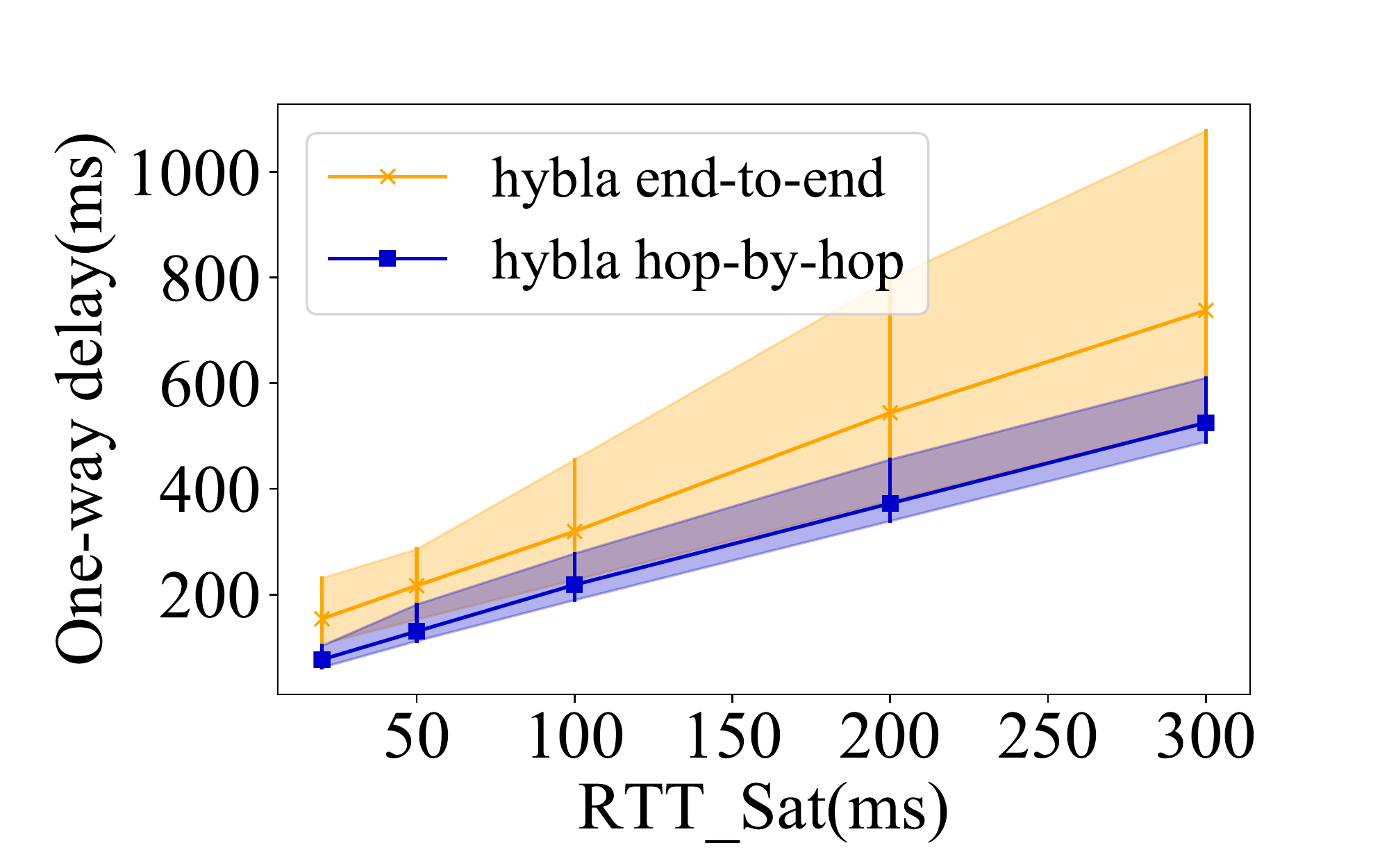}}
\label{fig_owd_hbhRetrans}
\end{minipage}
}
\caption{throughput and one-way delay plot in two-hop experiments.}
\label{fig_two_hop}
\end{figure*}

\begin{figure*}[!t]
\centering
\subfigure[CDF of one-way delay]{
\begin{minipage}[t]{0.28\linewidth}
\centerline{\includegraphics[width=1\linewidth]{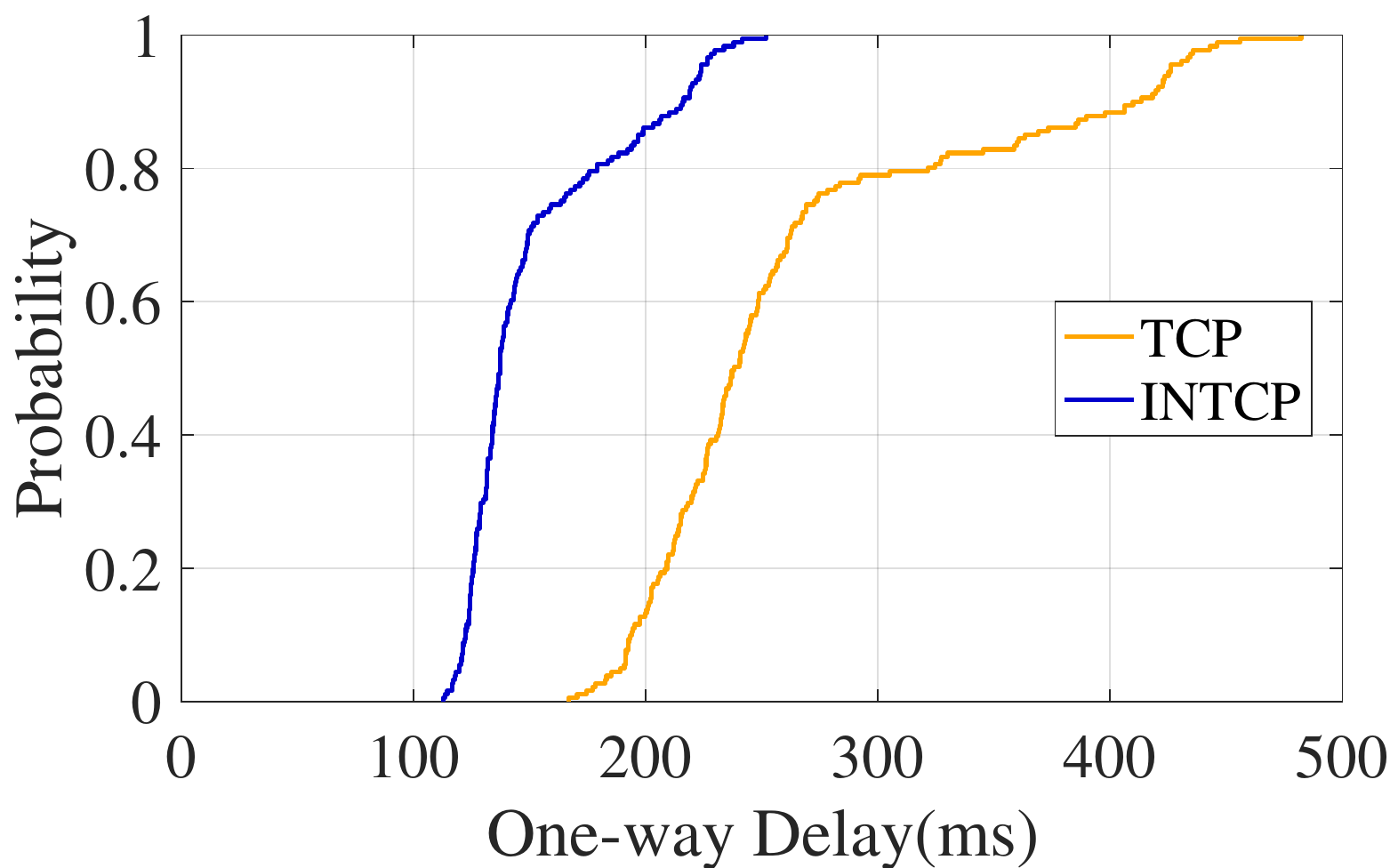}}
\label{fig_Unicast_owdCDF}
\end{minipage}
}
\subfigure[CDF of one-way delay jitter]{
\begin{minipage}[t]{0.28\linewidth}
\centerline{\includegraphics[width=1\linewidth]{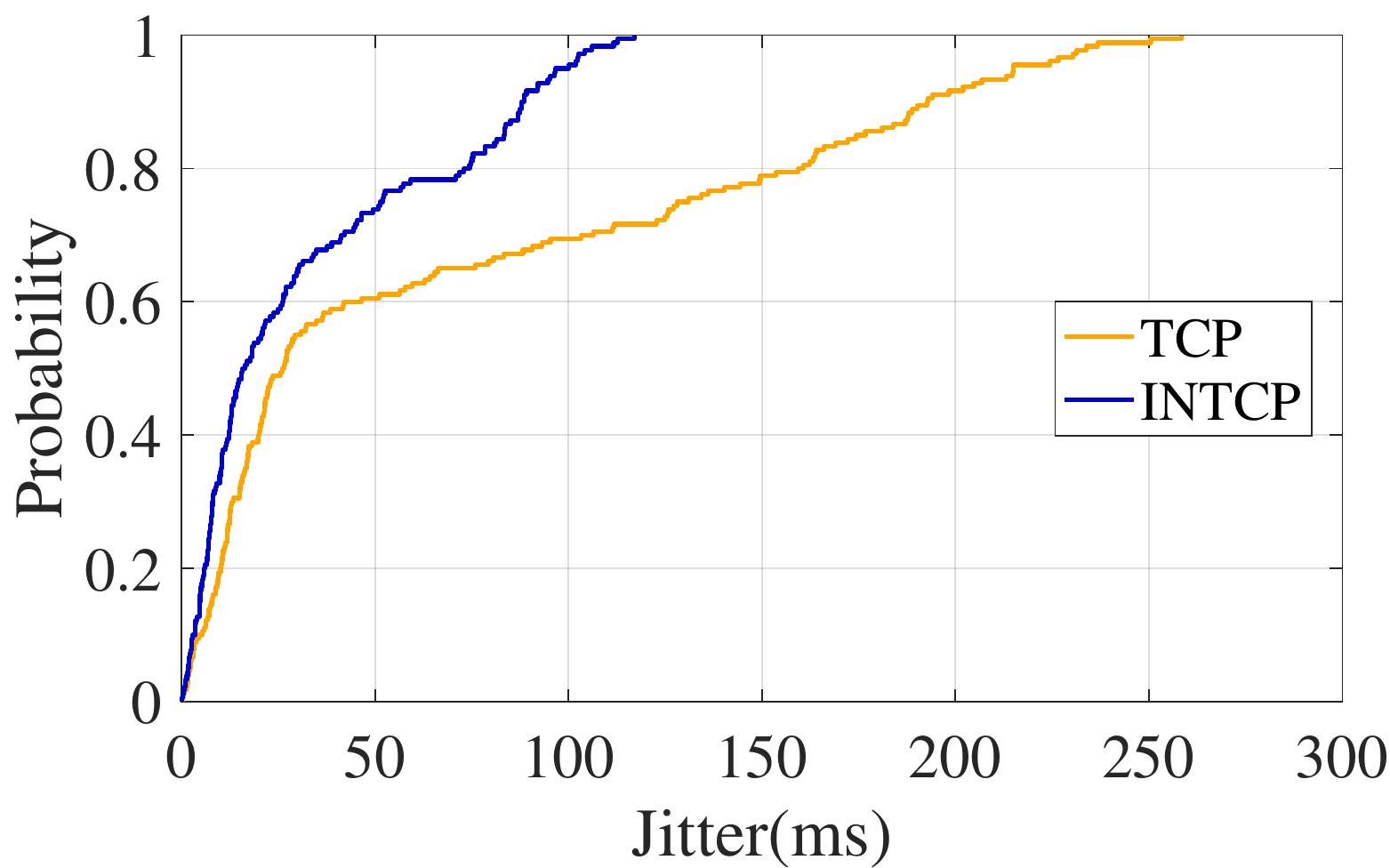}}
\label{fig_Unicast_jitterCDF}
\end{minipage}
}
\subfigure[CDF of throughput]{
\begin{minipage}[t]{0.28\linewidth}
\centerline{\includegraphics[width=1\linewidth]{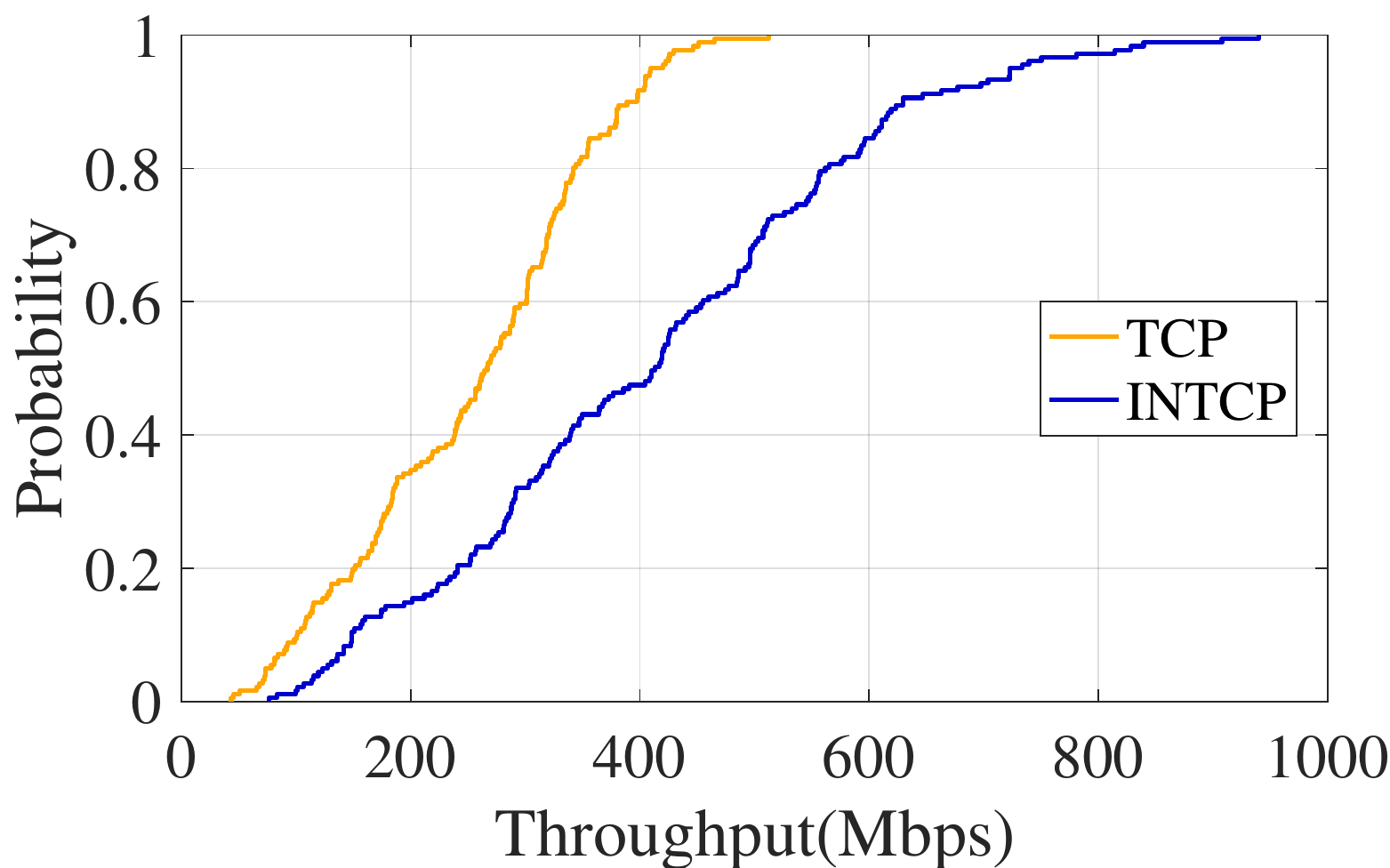}}
\label{fig_Unicast_thrpCDF}
\end{minipage}
}
\caption{Cumulative distribution function (CDF) plots of unicast experiments.}
\label{fig_Unicast_CDF}
\end{figure*}

\section{Performance Evaluations}
\label{sec_eva}
In this section, we evaluated INTCP in both unicast and multicast transmission scenes with the trace-based simulation.

\subsection{Methodology}
\textbf{Dataset:} Our experiment needs information of LEO links such as bandwidth, but there is currently no public dataset available, so we simulate them by building a satellite link model and use testing results published by Starlink~\cite{SpaceX} to set the model parameters. The core constellation of Starlink consists of 1,600 satellites, which are evenly distributed on 32 orbital planes at an altitude of 1150km with an inclination of 53 degrees to the equator. Based on this information, we calculate the satellites' physical coordinates at any time and use models such as Shannon formula to calculate the maximum physical bandwidth, packet loss rate, and delay of each link. Finally, we use the classic Dijkstra algorithm to obtain the routing table, thereby obtaining the dataset of testbed for the transport layer experiment.

\textbf{Testbed:} A network with up to hundreds of nodes, dynamic topology and varying link properties is necessary for simulating LEO constellation. The legacy network simulators, such as ns-2 and ns-3~\cite{NS}, work badly on supporting such a highly self-defined network, as well as having validity issues. If we use one real terminal or virtual machine as one node in satellite network, either the cost will be too high to be afforded or the performance will be awful. We build the testbed based on Mininet~\cite{Mininet}, which creates virtual networks using process-based virtualization and network namespaces that are available in recent Linux kernels. In this way, our satellite network testbed can run on a PC with ordinary hardware. Another advantage of this testbed compared to network simulator is that it supports standard Linux network software, so our implementation of INTCP protocol can be migrated to real Linux devices at zero cost. We choose a Windows PC (Intel i7-8700 CPU, 64GB RAM) as the platform to run the simulation experiment.

\textbf{Experiment setting:} Firstly, a set of experiments in two-hop network are displayed to evaluate the potential improvement of INTCP. This environment is that two endpoints and one satellite router constitute two hops, one for ground link, one for ground-satellite link(GSL). And the GSL parameters are man-made to observe the influence of various factors on network performance. Then we chose transcontinental communication as the simulating scenario because long-distance (more than 1000 kilometers) communication is a typical application scene of satellite networks~\cite{Mark18}. In addition, long-distance communication faces more serious problems such as high latency and dynamic topology. Specifically, we chose Beijing and New York as the endpoints of transmission. Both the unicast(single user) scenario and the multicast(multiple users requesting the same content) scenario are illustrated below.

\textbf{Performance metrics:} In the first two experiments, we chose three metrics to evaluate the performance of the network: one-way delay, jitter, and throughput. The definition of one-way delay is the time from that the responder to the requester. The definition of jitter is the delay difference between two consecutive seconds. The definition of throughput is the total number of bits received by the requester in a unit of time. In multicast scenario, we focus on the throughput metric.

\subsection{Two-hop Experiment}
In fig.~\ref{fig_varIntv}, we assume that the bandwidth of GSL fluctuates between 5Mbps and 35Mbps every period in a square wave while the ground link keeps 20Mbps. In the absence of cache, the throughput of end-to-end TCP at every moment is equal to $min(ground link bandwidth, GSL bandwidth)$. In hop-by-hop case, the cache on the router can store the data from ground when GSL bandwidth is low, and sends them out when GSL bandwidth is high. So even when the GSL bandwidth is higher than ground link, the GSL bandwidth is still fully used, making the average throughput of hop-by-hop INTCP much higher than end-to-end TCP.

Fig.~\ref{fig_itmDown} compares the ability to alleviating the impact of intermittent connectivity in TCP and INTCP. We assume a GSL handover event happens in every 20s. As the intermittent time grows, end-to-end TCP apparently suffers performance deterioration, while INTCP can alleviate the negative impact to a negligible level.

Fig.~\ref{fig_owd_hbhRetrans} shows the gain of on-hop loss recovery on one-way delay. In this experiment, the RTT of ground link is 50ms, the RTT of GSL is from 20ms to 300ms, the loss rate of GSL is 5\%. The curve in the center of each band in the figure indicates mean one-way delay. The upper/lower bound of each band indicates the 5\% / 95\% cumulative distribution point respectively. INTCP reduces average one-way delay and makes the variance smaller, which is important for some applications, e.g., video conference.

\subsection{Unicast Scenario}
In the evaluation of unicast, we transfer data from Beijing to New York, and ran the experiments for 24 hours in the simulation environment. The baseline is the legacy TCP protocol using the Reno congestion control algorithm, called TCP below.

Fig.~\ref{fig_Unicast_owdCDF} shows the Cumulative Distribution Function(CDF) of one-way delay. TCP Reno is prone to large delay (e.g., 300ms), while 80\% delay of INTCP is less than 180ms. It is because in an environment with a high packet loss rate, The performance of TCP drops sharply, and the hop-by-hop retransmission mechanism of INTCP can effectively alleviate this problem.

Fig.~\ref{fig_Unicast_jitterCDF} shows that the jitter of INTCP is almost always less than 100ms, while the jitter of TCP is much larger, because TCP will encounter huge jitter if there is end-to-end retransmission.

Fig.~\ref{fig_Unicast_thrpCDF} shows that the bandwidth of INTCP is significantly better than TCP, because the loss-based congestion control mechanism of Reno wastes a large part of physical bandwidth in a high-delay high-loss-rate network. Its throughput is constrained by the capacity of bottleneck link. While INTCP maximizes each link's bandwidth utilization. Cache multiplexes these links to improve INTCP throughput.

\subsection{Multicast transmission evaluation}
In multicast transmission evaluation, we assume that lots of users are evenly distributed on the route from Beijing to New York, requesting the same data. The data source is New York. Their request time is very concentrated which means that cache missing due to cache replacement can be ignored.

\begin{figure}[!t]
\centerline{\includegraphics[width=0.7\linewidth]{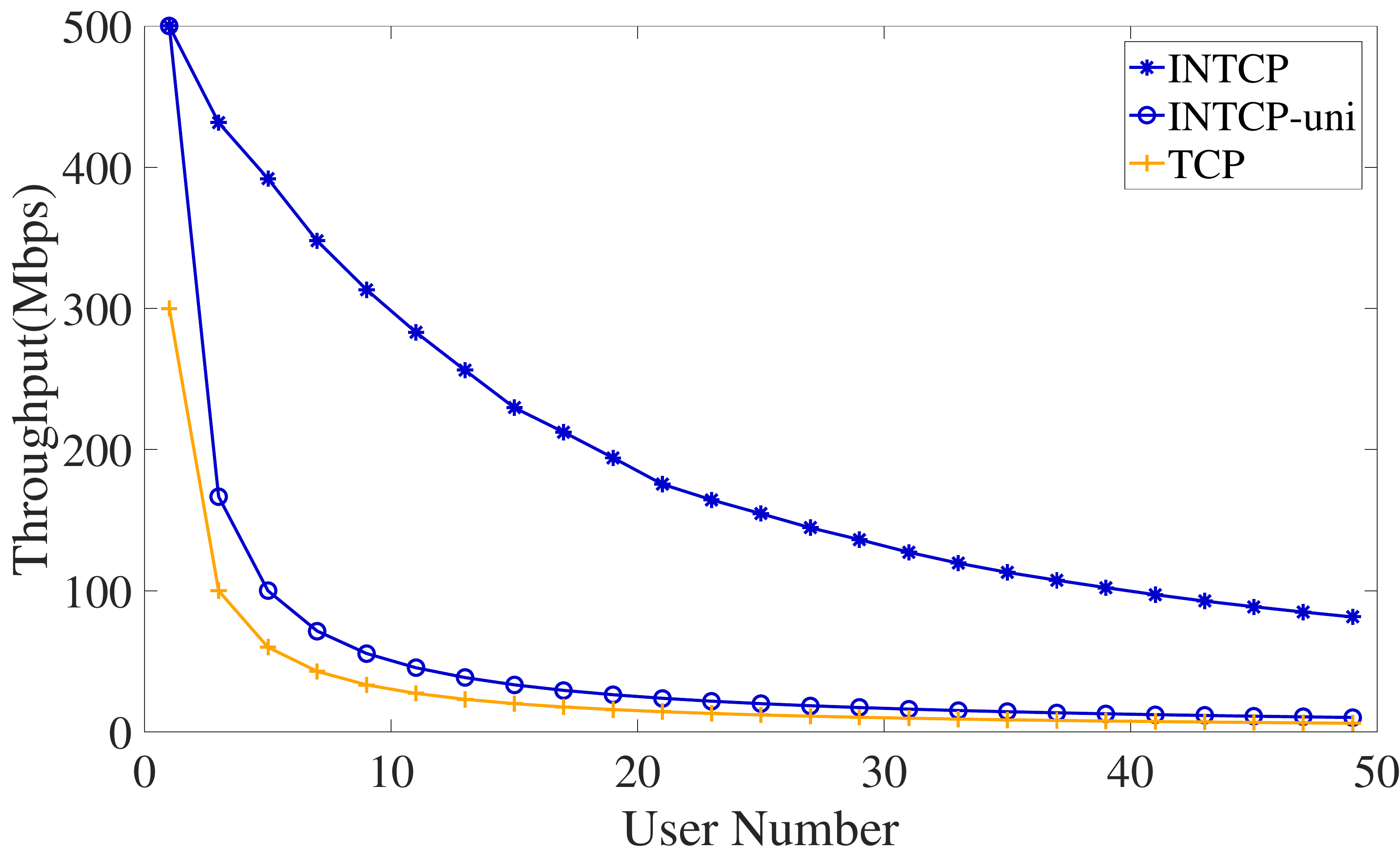}}
\caption{Average throughput VS. user number in multicast scenario.}
\label{fig_Multicast_thrpUserNum}
\end{figure}

In order to analyze the contribution of cache, we designed a variant of INTCP, called INTCP-uni, as the baseline. The difference of INTCP-uni is that it has no Cache. It retrievals data directly from the data source like legacy TCP.

Fig.~\ref{fig_Multicast_thrpUserNum} shows that as the number of users increases, the average throughput per INTCP-uni user drops sharply, while INTCP maintains a relatively high bandwidth. For the scenes such as live video streaming, this gap will bring users a big difference in the watching experience.


\section{Conclusion}
\label{sec_conclusion}
To provide high-speed and low latency Internet access on error-prone satellite links, it is necessary for satellite network to make great innovations on transport layer.
INTCP proposes an Information-centric hop-by-hop transport layer, which guarantees end-to-end reliable transmission by hop-by-hop loss recovery and congestion control. Hop-by-hop retransmission reduces the delay of end-to-end reliable transport, while hop-by-hop congestion control maximizes bandwidth utilization of links. Cache and Request-Response model are the basis of hop-by-hop transport layer. INCTP leverages them to recover lost packets and control flow's traffic per hop. The simulation experiments validate that INTCP improves average throughput about $60\%$, reduces delay $42\%$, jitter $53\%$. 
In the next step, the learning-based congestion control for satellite network will be investigated. The cost of running INTCP will be discussed and optimized by modifying the cache strategy and forwarding strategy.


%

\end{document}